\documentclass[a4paper,11pt]{article}
\usepackage{%
jcappub,
aas_macros,
braket,
}
\title{Constraining primordial vector mode from B-mode polarization}
\author[a]{Shohei Saga,}
\author[b,c]{Maresuke Shiraishi,}
\author[a,d]{Kiyotomo Ichiki}
\affiliation[a]{Department of Physics and Astrophysics, \\ 
Nagoya University, Aichi 464-8602, Japan}
\affiliation[b]{Dipartimento di Fisica e Astronomia ``G. Galilei'', \\ 
Universit\`a degli Studi di Padova, via Marzolo 8, I-35131, Padova, Italy}
\affiliation[c]{INFN, Sezione di Padova, \\ 
via Marzolo 8, I-35131, Padova, Italy}
\affiliation[d]{Kobayashi-Maskawa Institute for the Origin of Particles and the Universe, \\
Nagoya University, Nagoya 464-8602, Japan}

\abstract{The B-mode polarization spectrum of the Cosmic Microwave
Background (CMB) may be the smoking gun of not only the primordial tensor
mode but also of the primordial vector mode. If there
exist nonzero vector-mode metric perturbations in the early Universe,
they are known to be supported by anisotropic stress fluctuations of free-streaming
particles such as neutrinos, and to create characteristic signatures on
both the CMB temperature, E-mode, and B-mode polarization
anisotropies. We place constraints on the properties of the primordial
vector mode characterized by the vector-to-scalar ratio $r_{v}$ and the spectral
index $n_{v}$ of the vector-shear power
spectrum, from the {\it Planck} and BICEP2 B-mode data. We find that, for
scale-invariant initial spectra, the $\Lambda$CDM model including the
vector mode fits the data better than the model including the
tensor mode. The difference in $\chi^{2}$ between the vector and tensor
models is $\Delta\chi^{2} = 3.294$, because, on large scales the vector
mode generates smaller temperature fluctuations than the
tensor mode, which is preferred for the data.  In contrast, the
tensor mode can fit the data set equally well if we allow a significantly blue-tilted spectrum.  We
find that the best-fitting tensor mode has a large blue tilt and leads to
an indistinct reionization bump on larger angular scales.  The slightly red-tilted
vector mode supported by the current data set can also create ${\cal
O}(10^{-22})$-Gauss magnetic fields at cosmological recombination.  Our
constraints should motivate research that considers models of the early Universe that involve the vector mode.}

\emailAdd{saga.shohei@nagoya-u.jp}
\emailAdd{maresuke.shiraishi@pd.infn.it}
\emailAdd{ichiki@a.phys.nagoya-u.ac.jp}

\begin{document}

\maketitle
\flushbottom


\section{Introduction}

The Cosmic Microwave Background (CMB) radiation is an essential probe for verifying the standard cosmology and inflation models. 
The linear perturbations in the standard cosmology can be decomposed into the scalar mode (the curvature perturbation or the density perturbation) and the tensor mode (the primordial gravitational wave). 
Within these decompositions, the vector mode (i.e., the vorticity) is usually ignored since the vector mode only allows decaying solutions in the absence of sources.
However, there are many models containing non-decaying solutions in the vector-mode sector (e.g., refs.~\cite{Seljak:2006hi,Lewis:2004ef,Jacobson:2000xp}). 

The scalar mode is known to generate CMB temperature and E-mode polarization fluctuations. CMB observations by satellites, such as COBE \cite{Bennett:1996ce}, WMAP \cite{Bennett:2012zja}, and ${\it Planck}$ \cite{Ade:2013ktc,Ade:2013zuv} are consistent with theoretical predictions associated with the scalar-mode sector. These precise observations also determine the standard cosmological parameters and give us rich information about our Universe.

In addition to the temperature and E-mode fluctuations, the vector and
tensor modes can also induce B-mode fluctuations
\cite{Kamionkowski:1996zd,Seljak:1996gy,Zaldarriaga:1996xe}. Recently,
the BICEP2 experiment detected the B-mode
polarization with a tensor-to-scalar ratio normalized by $k = 0.05{\rm
Mpc^{-1}}$, $r_{0.05}=0.2^{+0.07}_{-0.05}$.
In other words, $r_{0.05}=0$ is
disfavored at 7$\sigma$ \cite{Ade:2014xna}.  The result may be attributed
straightforwardly to the detection of primordial
gravitational waves, which are well studied since they provide direct information on inflation in the early universe
\cite{1979JETPL..30..682S,1982PhLB..115..189R,Pritchard:2004qp}. The
primordial gravitational waves may explain the B-mode polarization
spectrum of the BICEP2 results for $\ell \lesssim 150$.  However, some debate exists over the BICEP2 result.  The
tensor-to-scalar ratio based on the BICEP2 B-mode data is somewhat
larger than the limit from the ${\it Planck}$ temperature data. Some works explain this discrepancy by introducing the
anti-correlated temperature spectrum
\cite{2014arXiv1403.5823K,Contaldi:2014zua,Kawasaki:2014fwa}. 
In addition, another issue discussed in the literature is on the excess of the B-mode
polarization spectrum at $150\lesssim \ell \lesssim 250$
\cite{Moss:2014cra,Lizarraga:2014eaa}. These studies suggest the
existence of something else, such as the vector mode.  After the release of
the BICEP2 results, some works discuss the constraints on
the vector mode induced by magnetic fields \cite{2014arXiv1403.6768B},
cosmic defects \cite{Moss:2014cra,Lizarraga:2014eaa} and self-ordering
scalar fields \cite{Durrer:2014raa}.  

Another solution to this discrepancy is foreground emission from the Galaxy. Indeed, it is
pointed out that the level of the foreground contamination in the B-mode
power spectrum presented in the
BICEP2 paper may be underestimated \cite{Mortonson:2014bja}, and the galactic,
polarized dust emission should contribute to some extent to the BICEP2 B-mode signal.
In ref.~\cite{Mortonson:2014bja} it is shown that
the foreground dust can mimic the polarization signal of the BICEP2
power spectrum and
can relax the above-mentioned tension between the temperature and polarization
anisotropies. However, at the moment we have no reliable
template for the foreground polarization emissions and must
wait for 
the {\it Planck} polarization result, which is expected to be available later
this year.

The purpose of the present paper is to update the constraints in Ref.~\cite{Ichiki:2011ah} including the B-mode polarizations and to investigate the role of B-mode polarizations in constraining the primordial tensor and vector modes.
 Although there are many scenarios including the vector mode, in the present paper, we focus in particular on the vector mode sustained by the anisotropic stress fluctuations of free-streaming particles such as neutrinos \cite{1992ApJ...392..385R,Lewis:2004kg,Ichiki:2011ah}. In the standard cosmology with perfect fluids, the vector mode contains only a decaying mode. However, the anisotropic stress fluctuations of free-streaming particles can lead to the non-decaying vector mode. It is known that this vector mode induces additional B-mode fluctuations and behaves quite differently from the primordial tensor mode.
The initial conditions of such a vector mode are characterized by two parameters associated with the vector-mode shear:
the vector-to-scalar ratio $r_{v}$ and the spectral index $n_{v}$ of the power spectrum. In the present paper, we show constraints on these parameters with the other relevant parameters from both the BICEP2 B-mode and {\it Planck} temperature data. 

This paper is organized as follows. In the next section, we briefly summarize the vector mode sustained by the anisotropic stress fluctuations of free-streaming particles and present the parameterization for the vector mode.
In section~\ref{sec:result}, we obtain constraints on the vector mode together with the other relevant parameters. The final section is devoted to the conclusion of the paper. 

\section{Vector mode induced by free-streaming particles}\label{sec:vector}

In this section, we review the vector mode induced by free-streaming particles \cite{Lewis:2004kg, Ichiki:2011ah}. The metric is given by
\begin{equation}
ds^{2}=a^{2}\left[ -d\eta^{2}+\left( \delta_{ij}+h_{ij}\right)dx^{i}dx^{j}\right] ~,
\end{equation}
where $\eta$ is the conformal time and $h_{ij}$ is the metric perturbation.
In the standard cosmological perturbation theory, the scalar, vector, and tensor modes do not mix in the evolution equations.
In this case, the metric perturbation of the vector mode can be written as
\begin{equation}
h_{ij}(\eta, {\bf x}) = \int{\frac{d^{3}k}{(2\pi)^{3}}}\sum_{\lambda =\pm 1} h^{(\lambda)} \left[ \hat{k}_{i}\epsilon^{(\lambda)}_{j}(\hat{k})+\hat{k}_{j}\epsilon^{(\lambda)}_{i}(\hat{k})\right]e^{i {\bf k} \cdot {\bf x}} ~,
\end{equation}
where we have translated in Fourier space. Here $\epsilon^{(\lambda)}_{i}(\hat{k})$ is the divergenceless polarization vector and $\lambda =\pm 1$ indicates the helicity state of the vector mode. Note that $h^{(+1)}$ and $h^{(-1)}$ evolve independently from each other because the usual Einstein gravity contains the parity symmetry.

From here on, we omit the superscript $\lambda$ for simplicity.
The Einstein equations for the vector mode are \cite{Lewis:2004kg, Ichiki:2011ah}
\begin{eqnarray}
\dot{\sigma}+2\mathcal{H}\sigma&=& -8\pi Ga^{2}p\Pi /k ~, \\
k^{2}\sigma&=&16\pi Ga^{2}q ~,
\end{eqnarray}
where we have defined the new variables, $\sigma\equiv \dot{h}/k$, $q \equiv (\rho +p)v$, and $\Pi$, which are the shear, the heat flux, and the anisotropic stress fluctuations for the vector mode, respectively.
Here a dot represents the derivative with respect to the conformal time.
Note that if the anisotropic stress fluctuations are not the source of the shear, the metric perturbation of the vector mode contains only a decaying mode. 

In the present analysis, we consider the standard cosmology with free-streaming particles, such as neutrinos. In this case, the vector shear $\sigma$ is effectively supported by nonzero anisotropic stress fluctuations of neutrinos posterior to their decoupling ($T \sim 1$ MeV) via the above Einstein equations. 
These variables act as sources of the CMB anisotropies during the recombination epoch \cite{Lewis:2004kg, Ichiki:2011ah}. 

\begin{figure}[t]
\begin{tabular}{cc}
\begin{minipage}{0.5\hsize}
\begin{center}
\rotatebox{0}{\includegraphics[width=1.0\textwidth]{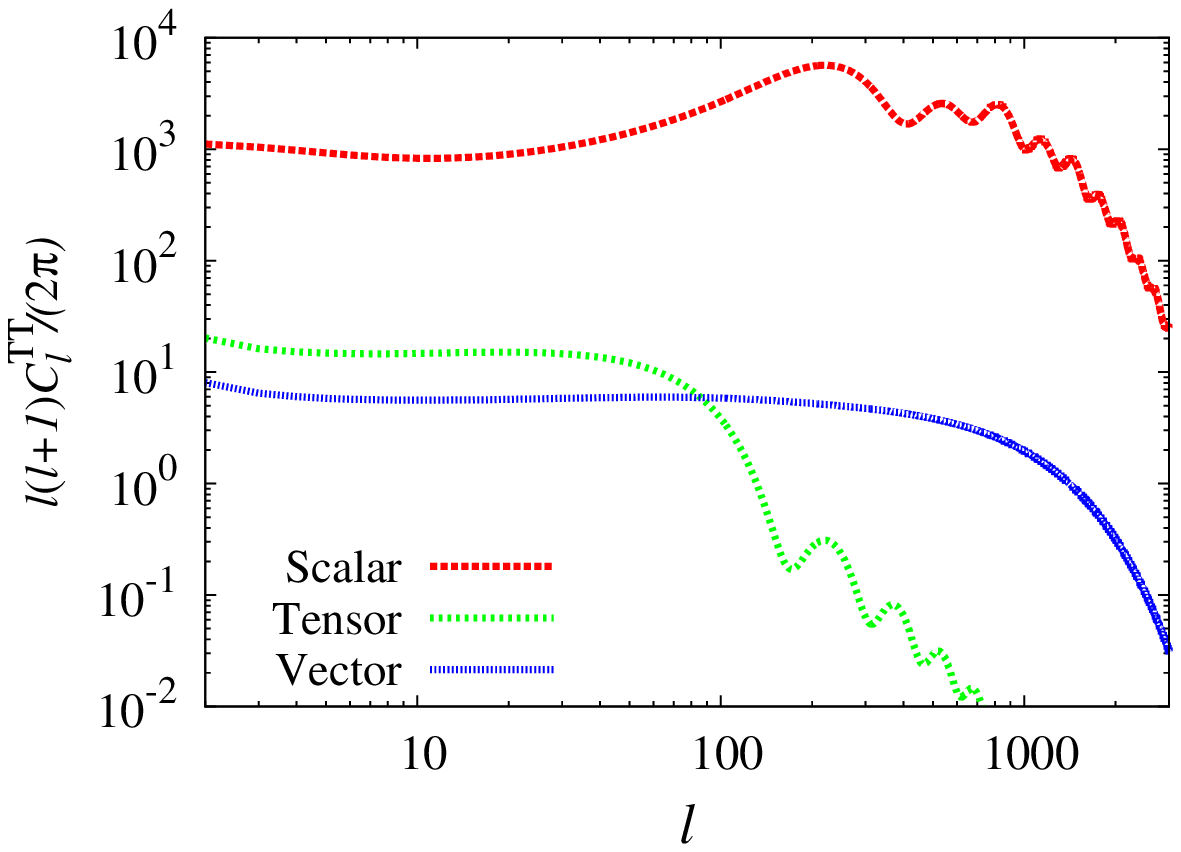}}
\end{center}
\end{minipage}
\begin{minipage}{0.5\hsize}
\begin{center}
\rotatebox{0}{\includegraphics[width=1.0\textwidth]{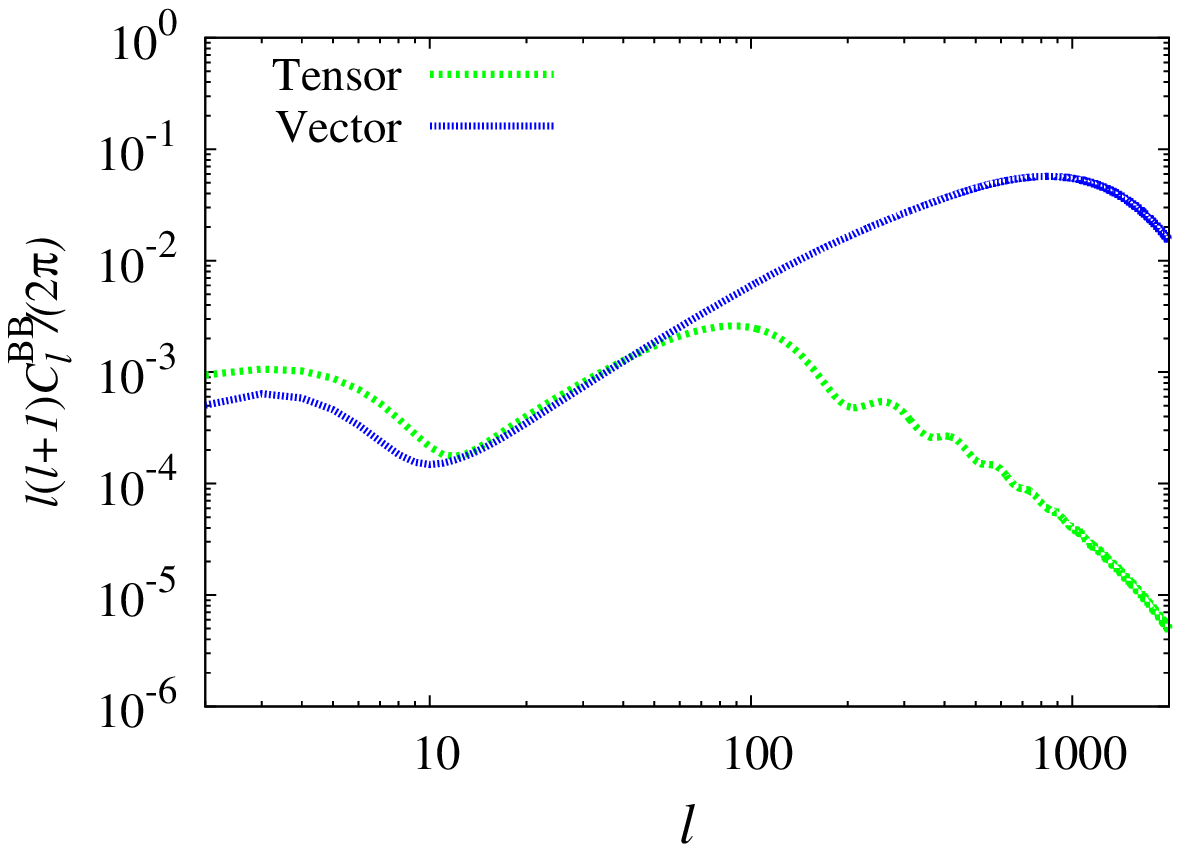}}
\end{center}
\end{minipage}
\end{tabular}
\caption{Power spectra of the CMB temperature (\textit{left}) and B-mode
 (\textit{right}) fluctuations induced from the scale-invariant vector
 ($n_v = 1$) and tensor $(n_t = 0)$ power spectra. For the vector
 mode, we use the best-fitting amplitude $r_{v} = 4.3 \times
 10^{-4}$ derived in section~\ref{sec:rv nv rt}, while the
 tensor-to-scalar ratio $r_{t}$ is adjusted to have the same B-mode amplitude at $50\lesssim \ell \lesssim 100$ for comparison.
}
\label{fig:CMB spectra ref}
\end{figure}

In figure~\ref{fig:CMB spectra ref}, we depict the CMB temperature fluctuation and B-mode polarization spectra for the scale-invariant cases, [i.e., $n_v = 1$ and $n_t = 0$ in eqs.~\eqref{eq:pow_vec} and \eqref{eq:pow_tens}], in order to show the shape of the spectra.
As shown in the figure, unlike for the tensor mode, the vector mode enhances small-scale temperature and polarization signals to the scale where the Silk damping is effective, namely $\ell \sim 1000$. More precisely, the temperature spectrum, $C^{TT}_{\ell}$ is mainly sourced by the baryon velocity in the Newtonian gauge, which is constant until the Silk damping effect occurs.
At this point, the temperature spectrum becomes constant for $\ell\lesssim 1000$.
However, the B-mode spectrum is sourced by the anisotropic stress fluctuations of photons and the E-mode polarization quadrupole.

Interestingly, because of these characteristics, if the vector and tensor B-mode spectra have almost the same amplitudes near $\ell= 100$, then the CMB temperature spectrum induced by the tensor mode is greater than that induced by the vector mode. This fact will yield a result that the vector mode is favored by the {\it Planck} and BICEP2 data in the scale-invariant case, which will be shown in section~\ref{subsec:one}.

Considering the vector mode quantitatively, we parameterize the primordial vector spectrum as follows
\begin{eqnarray}
\Braket{\sigma({\bf k})\sigma^{*}({\bf k'})} &\equiv& (2\pi)^{3}\frac{2\pi^{2}}{k^{3}}\mathcal{P}_{v}(k)\delta^{(3)}({\bf k}-{\bf k'}) ~, \\
\mathcal{P}_{v}&\equiv& \mathcal{A}_{v}\left( \frac{k}{k_{v0}}\right)^{n_{v}-1} ~, \label{eq:pow_vec}
\end{eqnarray}
where $\mathcal{A}_{v}$ is the amplitude of the primordial vector mode, $n_{v}$ is the spectrum index of the vector mode, $k_{v0} = 0.01~{\rm Mpc^{-1}}$ is the pivot scale of the vector mode, and the bracket means the ensemble average.
Furthermore, we define the vector-to-scalar ratio as
\begin{equation}
r_{v}\equiv \frac{\mathcal{A}_{v}}{\mathcal{A}_{s}} ~,
\end{equation}
where $\mathcal{A}_{s}$ is the primordial amplitude of the scalar mode. In terms of the power spectra of scalar and tensor modes, we follow the usual definitions: 
\begin{eqnarray}
\mathcal{P}_{s}(k) &=& \mathcal{A}_{s}\left( \frac{k}{k_{s0}}\right)^{n_{s}-1+\frac{1}{2}\alpha_{s}\ln{\left( k/k_{s0}\right)}} ~, \\
\mathcal{P}_{t}(k) &=& \mathcal{A}_{s}r_{t}\left( \frac{k}{k_{t0}}\right)^{n_{t}} ~, \label{eq:pow_tens}
\end{eqnarray}
where $r_{t}$, $n_{s}$, $n_{t}$, and $\alpha_{s}\equiv dn_{s}/d\ln{k}$ are the usual tensor-to-scalar ratio, the scalar spectral index, the tensor spectral index, and the scalar running index, respectively. In the present paper, we choose the pivot scales of the scalar and tensor modes to be $k_{s0} = 0.05~{\rm Mpc^{-1}}$ and $k_{t0} = 0.01~{\rm Mpc^{-1}}$, respectively.

\section{Constraints on the primordial vector mode}\label{sec:result}

In this section, we present constraints on the 
five parameters related to the vector and tensor modes and the scalar running index, i.e., $r_v$, $n_v$, $r_t$, $n_t$, and $\alpha_s$,
 from the {\it Planck} temperature, WMAP E-mode (WP), and BICEP2 B-mode
 data, combined with a compilation of Baryon Acoustic Oscillation (BAO) data
\cite{2011MNRAS.416.3017B,2012MNRAS.427.2132P,2012MNRAS.427.3435A}. The
 observed CMB power spectra are expressed as the sum of the scalar,
 vector, and tensor spectra as $C^{XX}_{\ell ~{\rm tot}}=C^{XX}_{\ell
 ~{\rm scal}}+C^{XX}_{\ell ~{\rm vec}}+C^{XX}_{\ell ~{\rm tens}}$, where
 $X=T$ (temperature), $E$ (E-mode polarization), and $B$ (B-mode
 polarization), respectively. We ignore the TB and EB correlations since
 there is no parity violation in the standard cosmology discussed in the
 present paper. 
To estimate the constraints on the vector mode, we
 modify the latest public 
 Monte-Carlo simulation code {\tt CosmoMC} \cite{Lewis:2002ah}, 
in which the BICEP2 likelihood files have already been included.

In the following analysis, we vary $r_{v}$, $n_{v}$, $r_{t}$, $n_{t}$
and $\alpha_{s}$ together with the default six parameters: the present baryon and CDM density parameters, the angular size of the sound horizon, the optical depth, and the amplitude and spectral index of the scalar power spectrum ($\Omega_{b}h^{2}$, $\Omega_{c}h^{2}$, $\theta_{\rm MC}$, $\tau$, $A_s$,~and $n_s$). The prior ranges are shown in Table~\ref{tab:prior}.
\begin{table}[t]
\begin{center}
\begin{minipage}{0.45\hsize}
\begin{center}
\begin{tabular}[t]{c|c}
\hline\hline
Parameter & Prior range \\
\hline
$\Omega_{b}h^{2}$ & [$0.005:0.1$] \\
$\Omega_{c}h^{2}$ & [$0.001:0.99$] \\
100$\theta_{\rm MC}$ & [$0.5:10$] \\
$\tau$ & [$0.01:0.8$] \\
$\ln{(10^{10}A_s)}$ & [$2.7:4.0$] \\
$n_{s}$ & [$0.9:1.1$] \\
\hline
\end{tabular}
\end{center}
\end{minipage}
\begin{minipage}{0.45\hsize}
\begin{center}
\begin{tabular}[t]{c|cc}
\hline\hline
Parameter & Prior range & Baseline \\
\hline
$\alpha_{s}$ & [$-0.5:0.5$] & $0$ \\ \hline
$r_{v}$ & [$0:0.2$] & $0$\\
$n_{v}$ & [$-2.0:6.0$] & $1$\\ \hline
$r_{t}$ & [$0:1.0$] & $0$\\
$n_{t}$ & [$-2.0:5.0$] & $0$ \\ \hline
\end{tabular}
\end{center}
\end{minipage}
\caption{Prior ranges of six standard parameters ({\it left}) and five additional ones ({\it right}).
The column ``Baseline'' gives the values used in the analysis when the corresponding parameter is fixed. Note that we do not use the consistency relation, $r_{t}=-8n_{t}$. In other words, the tensor spectral index $n_{t}$ runs freely.
The prior ranges for the vector mode parameters are matched to
 ref.~\cite{Ichiki:2011ah}.
The amplitudes of the CMB fluctuations induced by the vector mode are comparable to that by the tensor mode when $r_{v}/r_{t} \approx 10^{-2}$ \cite{Lewis:2004kg}.
Therefore we can set the vector-to-scalar ratio to be less than the tensor-to-scalar ratio.}
\label{tab:prior}
\end{center}
\end{table} 

Note that our prior ranges are chosen ad hoc, and not for a model selection through an evidence calculation. Instead, 
we focus on showing how the models fit the given data by comparing the likelihood or the effective chi-square values.
To decide which model is favored by given
data, one should account for the full prior range in the Bayesian
evidence.

\subsection{One-parameter estimate for scale-invariant cases}\label{subsec:one}

To begin with, let us estimate $r_v$ and $r_t$, separately with the assumption that the vector or tensor power spectrum is scale invariant, i.e. $n_v = 1$
or $n_t = 0$. The resulting effective chi-squares are 8717.774
and 8721.068 respectively for $r_{v}$ and $r_{t}$; the vector model provides a better fit to the
{\it Planck} temperature and BICEP2 B-mode data set than does
the tensor model.
This result occurs because
the amplitude of the
temperature power spectrum on large scales induced by the vector
mode is smaller than that by the tensor mode as shown in
figure~\ref{fig:CMB spectra ref}
when the vector or tensor mode with the scale
invariant spectra is fitted to the B-mode spectrum at the BICEP2
B-mode region, $50 \lesssim \ell \lesssim 300$. 
This result arises from the fact that, as we mentioned in the
previous section, the vector mode creates the CMB fluctuations on small
scales, while the tensor mode creates them on large scales.
In other words, the vector mode can create the B-mode polarization
with little effect on the temperature power spectrum compared
with the tensor one.

\subsection{Two-parameter estimate}

\begin{table}[t]
\begin{center}
\begin{minipage}{0.49\hsize}
\begin{center}
\begin{tabular}[t]{c|cc}
\hline\hline
\multicolumn{3}{c}{(i) Vector mode} \\
\hline
parameters & best fit & $68\%$ limits \\
\hline
$10^{4}r_{v}$ & $6.8$ & $6.2^{+1.14}_{-1.31}$ \\
$n_{v}$ & $0.47$ & $0.55^{+0.194}_{-0.263}$\\ \hline
\multicolumn{3}{c}{$\chi^{2}=8715.696$} $(\Delta\chi^{2} = 2.593)$\\
\hline
\end{tabular}
\end{center}
\begin{center}
\begin{tabular}[t]{c|cc}
\hline\hline
\multicolumn{3}{c}{(ii) Tensor mode} \\
\hline
parameters & best fit & $68\%$ limits \\
\hline
$r_{t}$ & $0.16$ & $0.17^{+0.045}_{-0.051}$ \\
$n_{t}$ & $2.0$ & $1.7^{+0.45}_{-0.40}$ \\ \hline
\multicolumn{3}{c}{$\chi^{2}=8713.103$} \\
\hline
\end{tabular}
\end{center}
\end{minipage}
\begin{minipage}{0.49\hsize}
\begin{center}
\begin{tabular}[t]{c|cc}
\hline\hline
\multicolumn{3}{c}{(iii) Running index} \\
\hline
parameters & best fit & $68\%$ limits \\
\hline
$r_{t}$ & $0.17$ & $0.19^{+0.037}_{-0.046}$ \\
$\alpha_{s}$ & $-0.029$ & $-0.029^{+0.0074}_{-0.0074}$ \\ \hline
\multicolumn{3}{c}{$\chi^{2}=8714.250$} $(\Delta\chi^{2} = 1.147)$\\
\hline
\end{tabular}
\end{center}
\begin{center}
\begin{tabular}[t]{c|cc}
\hline\hline
\multicolumn{3}{c}{(iv) Vector and Tensor modes} \\
\hline
parameters & best fit & $68\%$ limits \\
\hline
$r_{t}$ & $0.076$ & $0.069^{+0.032}_{-0.041}$ \\
$10^{4}r_{v}$ & $3.4$ & $4.1^{+1.3}_{-1.4}$ \\ \hline
\multicolumn{3}{c}{$\chi^{2}=8715.231$} $(\Delta\chi^{2} = 2.128)$ \\
\hline
\end{tabular}
\end{center}
\end{minipage}
\caption{Best-fit values and $1\sigma$ (marginalized) limits on model parameters in analyses with vector mode alone ($r_v, n_v$) (\textit{top left}), the tensor mode alone ($r_t, n_t$) (\textit{bottom left}), the BICEP2 like parameterization $(r_{t}, \alpha_{s}, n_{t}=0)$ (\textit{top right}) and the vector and tensor modes with scale invariant spectra $(r_{v}, r_t, n_v = 1, n_{t}=0)$ (\textit{bottom right}). 
Here, $\Delta\chi^{2}$ in the brackets gives the differences of $\chi^2$ from the value in the tensor mode alone parameterization.
The results show that $\chi^2$ for the tensor mode alone case is the smallest among the four cases.
}
\label{tab:constraints}
\end{center}
\end{table}

\begin{figure}[t]
\begin{tabular}{cc}
\begin{minipage}{0.5\hsize}
\begin{center}
\rotatebox{0}{\includegraphics[width=1.0\textwidth]{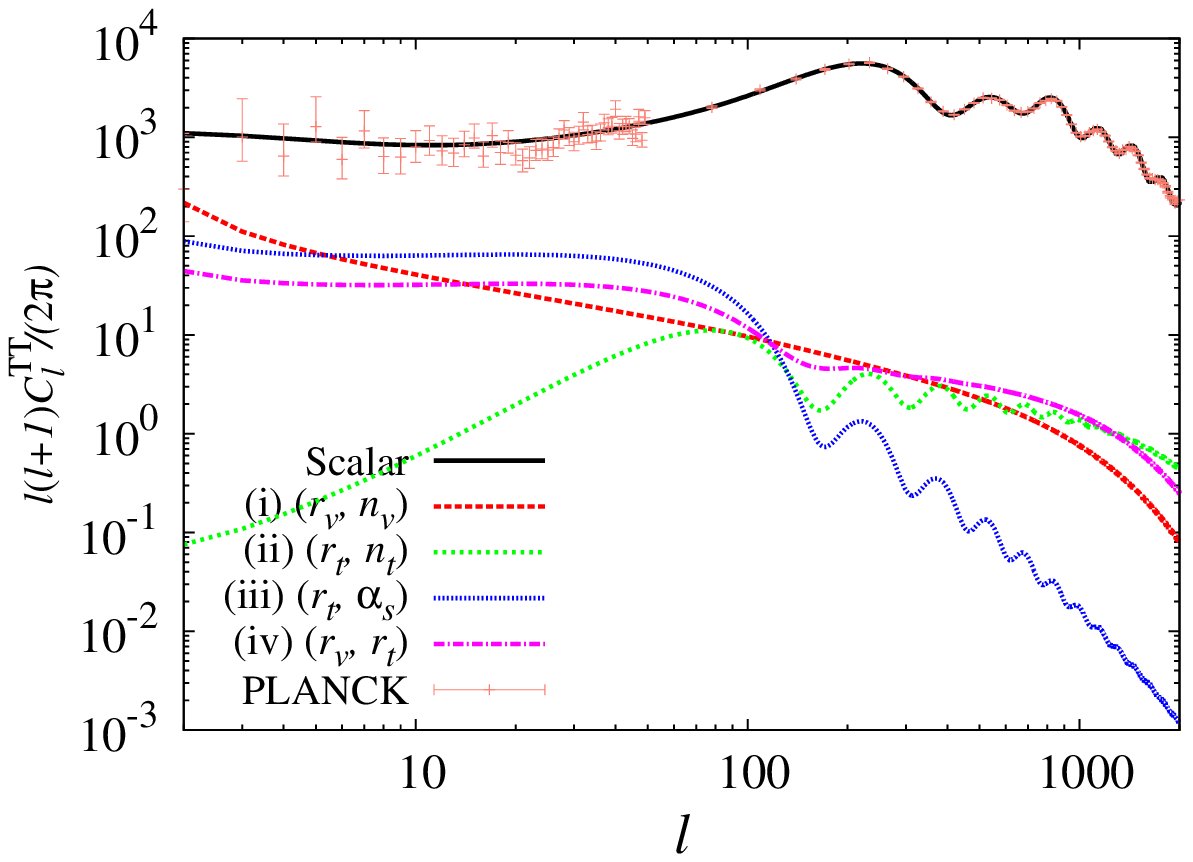}}
\end{center}
\end{minipage}
\begin{minipage}{0.5\hsize}
\begin{center}
\rotatebox{0}{\includegraphics[width=1.0\textwidth]{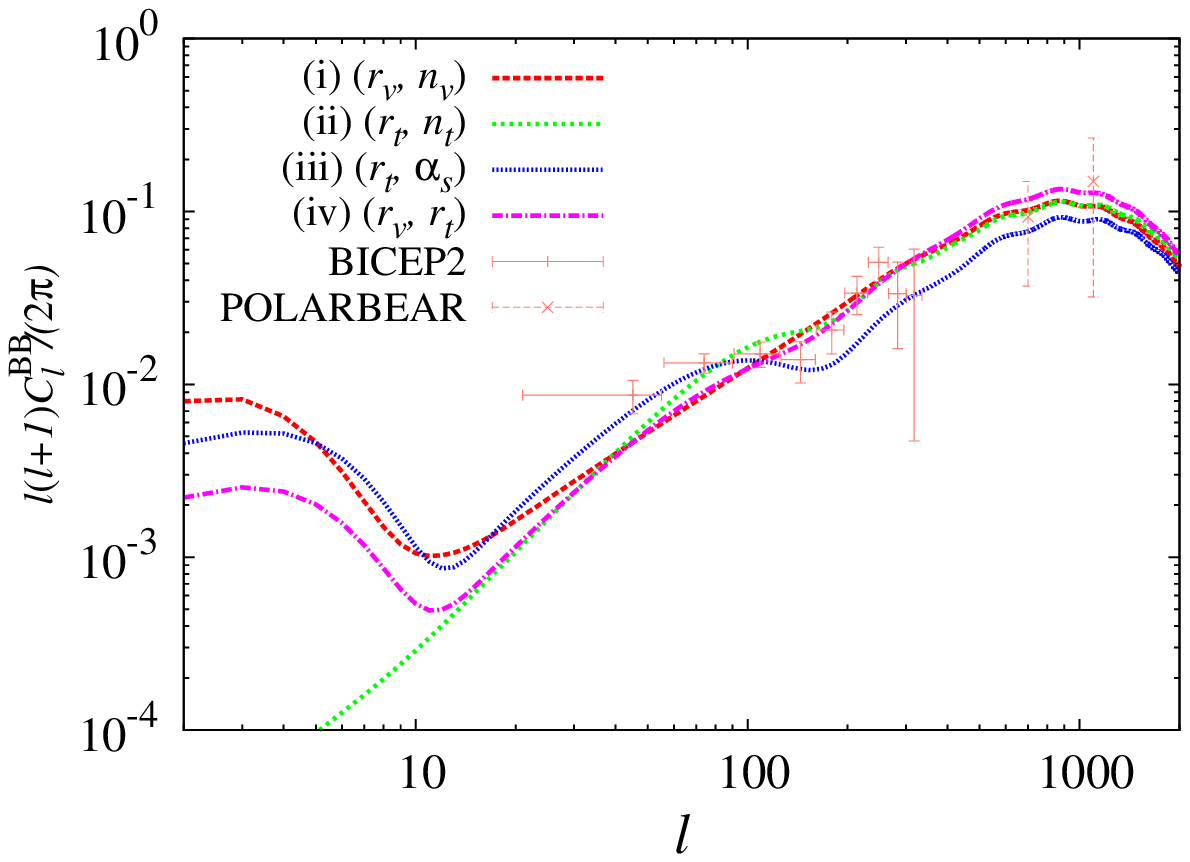}}
\end{center}
\end{minipage}
\end{tabular}
\caption{Power spectra of CMB temperature (\textit{left}) and B-mode (\textit{right}) fluctuations with best-fit parameters given in Table.~\ref{tab:constraints}.
Note that the bump of the reionization epoch in the case of $(r_{t}, n_{t})$ becomes ambiguous since this parameterization leads to a very blue spectrum to suppress on large scales in the temperature spectrum \cite{Wang:2014kqa,Gerbino:2014eqa,Liu:2014tda}.
For reference, the spectrum induced by the adiabatic scalar-mode is also
 shown in the left panel.
We also show the B-mode spectrum detected by the POLARBEAR experiment \cite{Ade:2014afa}.
We do not use the POLARBEAR data in our analysis as they are not as constraining as the other datasets considered.}
\label{fig:CMB spectra}
\end{figure}

We now constrain the four sets of the two model parameters: (i)
$(r_{v}, n_{v})$, (ii) $(r_{t}, n_{t})$, (iii) $(r_{t}, \alpha_{s})$ and
(iv) $(r_{v}, r_{t})$. The results are shown in
Table~\ref{tab:constraints}. We find that the ($r_{t}, n_{t}$) case can
minimize the effective chi-square; namely, the tensor mode-only case with
the relaxed inflationary-consistency relation fits the
data better than the vector-mode cases ($r_{v}, n_{v}$)  and ($r_{t},
\alpha_{s}$) do.

The reasons for this result may come from analyzing figure~\ref{fig:CMB spectra}, which shows the best-fit CMB temperature and B-mode power spectra.
The right panel shows that the four curves are very nearly consistent with each other for $\ell > 10$ and all of them fit the BICEP2 B-mode data well. However, as seen in the left panel, the shape of each temperature spectrum is quite different. The case of $(r_t, n_t)$ is consistent with the BICEP2 data when the tensor mode spectrum is blue-tilted, i.e., $n_t = 2.0$. In this case, the large-scale temperature signals are damped and are consistent with the {\it Planck} temperature data \cite{Wang:2014kqa,Gerbino:2014eqa,Liu:2014tda}. 
Furthermore, damping the large-scale B-mode signals makes the reionization bump ambiguous.
However the other three cases are less suitable because of the large-scale temperature contributions. In particular, the temperature signals from the vector-mode case $(r_v, n_v)$ are significant on very large scales because a slightly red-tilted spectrum ($n_v = 0.47$) is required to fit the BICEP2 B-mode data, making it most inconsistent.

\subsection{Constraints on $r_{v}$, $n_{v}$, and $r_{t}$}\label{sec:rv nv rt}

In this subsection, we constrain the three additional parameters $r_{v}$, $n_{v}$, and $r_{t}$ at the same time, as discussed in ref.~\cite{Ichiki:2011ah}. In that paper, the authors used the observational data of the temperature and E-mode polarization given only by the WMAP experiment. We show here how the constraints are improved by the new temperature and B-mode data from ${\it Planck}$ and BICEP2. 
Note that the running index of the initial scalar amplitude $\alpha_{s}$ is fixed to zero in this subsection.

\begin{figure}[t]
\begin{tabular}{cc}
\begin{minipage}{0.5\hsize}
\begin{center}
\rotatebox{0}{\includegraphics[width=1.0\textwidth]{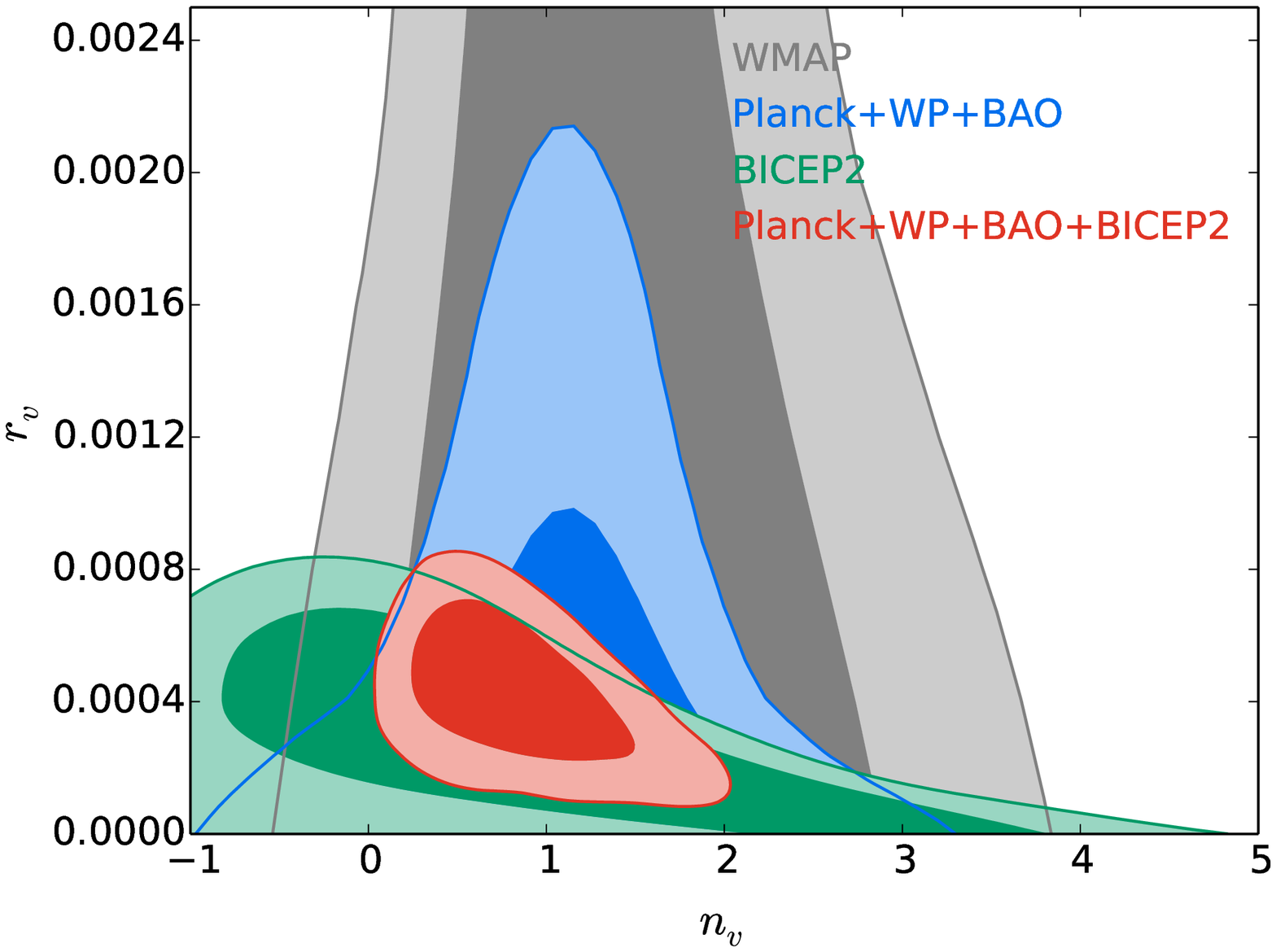}}
\end{center}
\end{minipage}
\begin{minipage}{0.5\hsize}
\begin{center}
\rotatebox{0}{\includegraphics[width=1.0\textwidth]{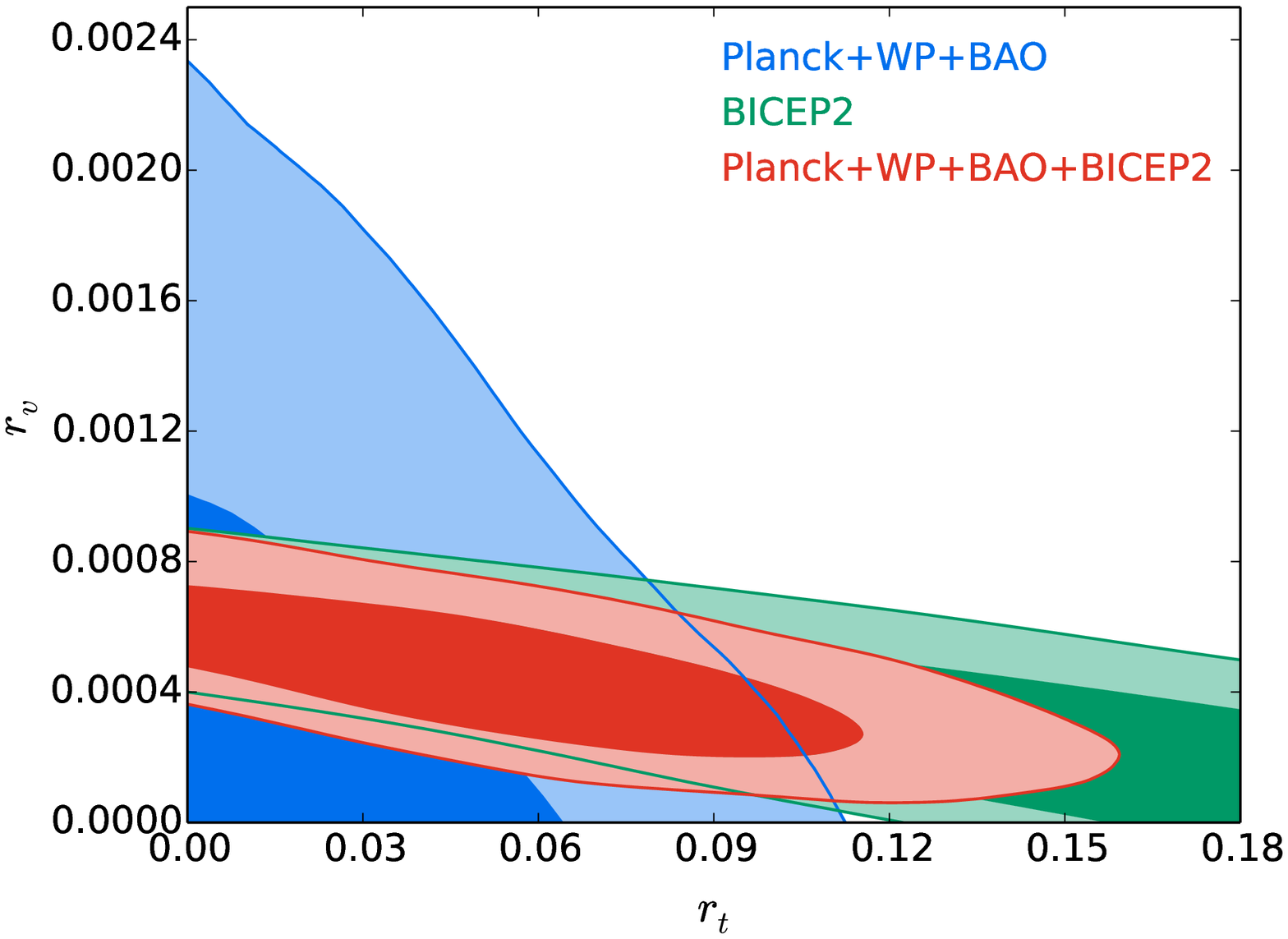}}
\end{center}
\end{minipage}
\end{tabular}
\caption{Marginalized two-dimensional posteriors for ($r_{v}, n_{v}$)
(\textit{left}) and ($r_{v}, r_{t}$) (\textit{right}) obtained from {\it
Planck}+WP (blue) and {\it Planck}+WP+BICEP2 (red). To compare with
the previous study \cite{Ichiki:2011ah}, in the {\it left} panel, we
also describe the constraint from the WMAP temperature and polarization
data (gray).  For reference, we also show the marginalized
two-dimensional posteriors obtained by the BICEP2 data alone (green)
which are derived with the six standard cosmological parameters
fixed to the {\it Planck} results as in
Ref.~\cite{Gerbino:2014eqa}.  } \label{fig:contour}
\end{figure}

Figure~\ref{fig:contour} depicts the marginalized two-dimensional posteriors for ($r_{v}, n_{v}$) and ($r_{v}, r_{t}$).
Note that we assume $n_{t} = 0$, (i.e., the scale-invariant spectrum of the tensor mode), instead of the usual consistency relation $r_{t} = -8 n_{t}$. 
The figure shows that the {\it Planck} temperature and BICEP2 B-mode data constrain $r_{t}$, $r_{v}$, and $n_{v}$ strongly.
From the {\it left} panel, we notice the interesting result that the BICEP2 data fits a smaller $n_{v}$ whereas the other data without the B-mode indicates a somewhat larger $n_{v}$. 
Concerning $r_{v}$ and $r_{t}$, the {\it right} panel tells us that, by combining the BICEP2 result with the other data, a nonzero $r_{v}$ is favored, while $r_{t}$ is still consistent with null. 
Again, this is because the vector mode does not modify the temperature anisotropies as much as does the tensor mode for the nearly scale-invariant cases if the magnitude of the resulting B-mode spectrum is comparable to the BICEP2 data, as discussed in section~\ref{sec:vector}. The marginalized limits for the {\it Planck}+WP+BAO+BICEP2 estimate are, given by 
$r_{v} = 4.3^{+1.6}_{-1.6} \times 10^{-4}$, 
$n_{v} = 0.83^{+0.28}_{-0.48}$, and 
$r_{t} < 0.09$.

In ref.~\cite{Ichiki:2011ah}, the authors also discuss the possibility
that this vector mode creates seed magnetic fields that
may be amplified through the dynamo mechanism to become the micro-Gauss-level 
magnetic fields associated with galaxies and clusters of galaxies (e.g. ref.~\cite{Widrow:2011hs}). 
The dynamo process needs the seed fields whose strength are as large as $10^{-20} \sim 10^{-30} {\rm Gauss}$ at the recombination epoch \cite{Davis:1999bt}.
Theoretically, the vorticity driven by the vector shear produces
rotational electric fields that can change to magnetic fields via the
Maxwell constraint equation. The authors showed that ${\cal
O}(10^{-21})$-Gauss magnetic fields at $k\approx 0.1~{\rm Mpc}^{-1}$,
which are required at the recombination epoch, can be produced by the
$r_{v}$ and $n_{v}$ allowed by WMAP. Here, let us consider the generation of magnetic fields with the
new parameter regions. When we adopt the above best-fit parameters, the
amplitude of magnetic fields is about $10^{-22}$ Gauss at $k = 0.1~{\rm
Mpc}^{-1}$ at cosmological recombination. This value is comparable to or
slightly less than the result of ref.~\cite{Ichiki:2011ah} but
the amplitude may be large enough to be the seed magnetic fields.

\section{Conclusion}

If vector modes exist in the early Universe, and they evolve with the support of anisotropic stress fluctuations of free-streaming particles like neutrinos, they produce CMB temperature, E-mode, and B-mode anisotropies whose shapes are quite different from the scalar-mode and tensor-mode anisotropies typically considered.
In the present paper, we investigate how such vector-mode signals are constrained from the latest temperature and B-mode data provided by {\it Planck} and BICEP2 in comparison with the tensor mode.
Next, by using a Monte-Carlo approach, we estimate observational bounds on the vector-to-scalar ratio, $r_{v}$, and the spectral index, $n_{v}$, associated with the primordial power spectrum of the vector-mode shear, combined with the usual scalar-mode and several tensor-mode parameters.

Upon estimating the parameter from the scale-invariant spectra, the interesting result is that the vector mode creates the B-mode polarization without modifying the large-scale temperature spectrum, compared with the tensor mode.
When varying the spectral indices, in which the tensor power spectrum may be tilted, the tensor mode with a large blue tilt fits both the {\it Planck} and the BICEP2 data.
This is because the large-scale temperature spectrum induced by the tensor mode can be suppressed and
the tensor mode does not destroy the scalar mode fit to the large scale temperature data from {\it Planck}.
In this case, we should note that the large-scale B-mode polarization is also suppressed and the bump from the reionization epoch vanishes.
Thus, a precise observation of the reionization bump is important.
The best chance upcoming experiments will have of revealing the source of the B-mode polarization spectrum is by observing the spectrum on large scales, as experiments aiming at smaller scales must differentiate between the primordial signal and the lensing B-mode contaminant. Precise observations at large scales could clarify the source of the primordial B-mode spectrum: blue tensor modes, ``standard'' tensor modes or vector modes.

The vector mode can also produce magnetic fields in the primordial photon and baryon plasma. The parameter values allowed by the {\it Planck} and BICEP2 data set indicate that the strength of the resulting magnetic fields is about $10^{-22}$~Gauss at $k = 0.1~{\rm Mpc}^{-1}$ at cosmological recombination. This value is comparable to or slightly less than that needed to explain the observed large-scale magnetic fields; namely, ${\cal O}(10^{-21})$~Gauss.

The current CMB data are well described by the existence of
slightly red primordial vector mode, which may play a role in
generating cosmic magnetic fields. 
Probing early-Universe models that produce such vector modes is an exciting avenue for future research.

\acknowledgments
This work was supported in part by a Grant-in-Aid for JSPS Research under Grants Nos.~26-63~(SS) and 25-573~(MS), and a Grant-in-Aid for Scientific Research under No.~24340048 (KI). M.S. was also supported in part by the ASI/INAF Agreement I/072/09/0 for the Planck LFI Activity of Phase E2. We also acknowledge
the Kobayashi-Maskawa Institute for the Origin of Particles and the Universe, Nagoya University, for providing computing resources useful in conducting the research reported in this paper. This research has also been supported in part by World Premier International Research Center Initiative, the Ministry of Education, Culture, Sports, Science and Technology, Japan.



\bibliography{paper}
\end{document}